\documentclass[12pt,BCOR0mm,DIV13]{scrartcl}

\usepackage{amsmath,amsfonts,typearea,booktabs,graphicx}
\usepackage{xcomment,color,xspace,hyperref}

\usepackage{harvard}
\usepackage[british]{babel}
\usepackage{microtype}
\graphicspath{ {graphics/} }
\newtheorem{proposition}{Proposition}
\newcommand{\bmb}{\mathbf{b}}
\newcommand{\bmc}{\mathbf{c}}
\newcommand{\bmA}{\mathbf{A}}
\newcommand{\bmX}{\mathbf{X}}
\newcommand{\bmR}{\mathbf{R}}
\newcommand{\bmU}{\mathbf{U}}
\newcommand{\bmh}{\mathbf{h}}
\newcommand{\bmx}{\mathbf{x}}
\newcommand{\bmm}{\mathbf{m}}
\newcommand{\bmW}{\mathbf{W}}
\newcommand{\bmG}{\mathbf{G}}
\newcommand{\bmI}{\mathbf{I}}
\newcommand{\bmM}{\mathbf{M}}
\newcommand{\bmF}{\mathbf{F}}
\newcommand{\bmV}{\mathbf{V}}
\newcommand{\bmY}{\mathbf{Y}}
\newcommand{\bmy}{\mathbf{y}}
\newcommand{\bmu}{\mathbf{u}}

\newcommand{\bmalpha}{\mbox{\boldmath$\alpha$}}
\newcommand{\bmbeta}{\mbox{\boldmath$\beta$}}
\newcommand{\bmeta}{\mbox{\boldmath$\eta$}}
\newcommand{\bmPsi}{\mbox{\boldmath$\Psi$}}
\newcommand{\bmzero}{\mathbf{0}}

\newcommand{\oftime}[1]{({#1})}
\newcommand{\ofr}{(r)}
\newcommand{\oft}{(t)}

\newcommand{\Prob}[1]{\mathbb{P} \left({#1}\right)}
\newcommand{\Norm}[1]{\left\vert\left\vert{#1}\right\vert\right\vert}
\newcommand{\Abs}[1]{\left\vert{#1}\right\vert}

\begin{document}

\title{Bayesian Inference for Hybrid Discrete-Continuous Stochastic Kinetic
  Models}
\author{Chris Sherlock$^1$, Andrew Golightly$^2$\footnote{andrew.golightly@ncl.ac.uk} ~and Colin S. Gillespie$^2$}
\date{{\small $^1$Department of Mathematics and Statistics, Lancaster University, UK\\
$^{2}$School of Mathematics \& Statistics, Newcastle University, UK \\}}

\maketitle

\begin{abstract}
  \noindent We consider the problem of efficiently performing simulation and
  inference for stochastic kinetic models. Whilst it is possible to work
  directly with the resulting Markov jump process, computational cost can be
  prohibitive for networks of realistic size and complexity. In this paper, we
  consider an inference scheme based on a novel hybrid simulator that classifies
  reactions as either ``fast'' or ``slow'' with fast reactions evolving as a
  continuous Markov process whilst the remaining slow reaction occurrences are
  modelled through a Markov jump process with time dependent hazards. A linear
  noise approximation (LNA) of fast reaction dynamics is employed and slow
  reaction events are captured by exploiting the ability to solve the stochastic
  differential equation driving the LNA. This simulation procedure is used as a
  proposal mechanism inside a particle MCMC scheme, thus allowing Bayesian
  inference for the model parameters. We apply the scheme to a simple
  application and compare the output with an existing hybrid approach and also a
  scheme for performing inference for the underlying discrete stochastic model.
  \medskip

\noindent \textbf{Keywords:} Stochastic kinetic model, linear noise
approximation, Poisson thinning, particle MCMC
\end{abstract}

\section{Introduction}

A growing realisation of the importance of stochasticity in cell and molecular
processes \cite[for example]{mcadams1999,kitano2001,swain2002} has stimulated the need for efficient
methods of inferring rate constants in stochastic kinetic models (SKMs) associated with
gene regulatory networks. Such inferences are typically required to allow
predictive \textit{in silico} experiments. Performing inference for the Markov jump 
process representation of the SKM is straightforward given observations on all
reaction times and types. In this case, it is possible to construct a complete 
data likelihood, for which a conjugate analysis is possible \cite{wilkinson2012}. 
In practice, a subset of species may be observed at discrete times.
\citeasnoun{boys2008} show that it is possible to construct 
Metropolis-Hastings schemes for performing inference in this setting. 
However, the statistical efficiency of such schemes can be poor, and these
methods are likely to be more computationally demanding than simulating the
process exactly (using, for example, the Gillespie algorithm \cite{gillespie1977}). 
Therefore, whilst inference in this setting is possible in theory, 
in practice computational cost precludes analysis of systems of realistic size.

Considerable speed-up can be obtained by ignoring discreteness and stochasticity
in the inferential model. For example, the macroscopic rate equation (MRE)
models the dynamics with a set of coupled ordinary differential equations
\cite{kampen2001}. Computational savings can still be made when
  adopting the diffusion approximation or chemical Langevin equation
(CLE) \cite{Gillespie2000} on the other hand, which ignores discreteness but not
stochasticity by modelling the biochemical network with a set of coupled
stochastic differential equations (SDEs). Although the transition density
characterising the process under the CLE is typically intractable, it has been
shown that basing inference algorithms around this model can work well for some
applications \cite{golightly2005,Heron07,Purutcuoglu07,golightly11,picchini13}. Further
computational gains can be made by adopting a linear noise approximation (LNA)
of the CLE \cite[for example]{kampen2001} which is given by the MRE plus a
stochastic term accounting for random fluctuations about the MRE. Under the LNA,
the transition density is a tractable Gaussian density (provided that the
initial value is fixed or follows a Gaussian distribution).
Performing inference for the LNA has been the focus of
\citeasnoun{Komorowski09}, \citeasnoun{stathopoulos13} and
\citeasnoun{sherlock2012} among others. However, biochemical reactions
describing processes such as gene regulation can involve very low concentrations
of reactants \cite{guptasarma1995} and ignoring the inherent discreteness in low
copy number data traces is clearly unsatisfactory.

The aim of this paper is to exploit the computational efficiency of methods such
as the CLE and LNA whilst accurately describing the dynamics of low copy number
species. Hybrid strategies for simulating from discrete-continuous stochastic kinetic
models are reasonably well developed and involve partitioning reactions as fast
or slow based on the likely number of occurrences of each reaction over a given
time interval and the effect of each reaction on the number of reactants and
products. Use of the CLE to model fast reaction dynamics in order
to simulate efficiently from an approximation to the system has been the focus of
\citeasnoun{haseltine02}, \citeasnoun{burrage04}, \citeasnoun{salis2005} and
\citeasnoun{higham2011} amongst others. Discrete/ODE approaches (e.g.
\citeasnoun{kiehl2004} and \citeasnoun{alfonsi2005}) are also possible and we
refer the reader to \citeasnoun{pahle09} and \citeasnoun{golightly2013} for
recent reviews. Since the slow reaction hazards will necessarily depend on
species involved in fast reactions, these hazards are typically not constant
between slow reaction events, and efficient sampling of these slow event times
can be problematic.

We propose a novel hybrid simulation strategy that models fast reaction dynamics
with the LNA and slow dynamics with a Markov jump process. Moreover, by deriving
a probable upper bound for a combination of components that drive the LNA, we
obtain a probable upper bound for the total slow reaction hazard. This allows
efficient sampling of the slow reaction times via \textit{thinning}, which is a
point process variant of rejection sampling \cite{lewis1979}. Related approaches
have been proposed by \citeasnoun{casella2011} and \citeasnoun{rao2013}. The
former consider simulation for jump-diffusion processes by combining a thinning
algorithm with a generalisation of the exact algorithm (for diffusions)
developed by \citeasnoun{beskos2005}, whilst the latter assume that an upper
bound for the rate matrix governing the MJP is available and use uniformisation
\cite{hobolth2009} to simulate the process.

We use our approximate model to perform Bayesian inference for the governing
kinetic rate constants using noisy data observed at discrete time points. In
particular, we focus on a special case of the particle marginal Metropolis
Hastings (PMMH) algorithm \cite{andrieu2010} which targets the marginal
posterior density of the model parameters and permits exact, simulation-based
inference. The algorithm requires implementation of a particle filter
\cite{carpenter1999,pitt1999,doucet2000,delmoral2002} in the latter step, and we
apply the bootstrap filter \cite{gordon1993} which only requires the ability to
forward simulate from the model and evaluate the observation densities
associated with each data point. Use of our novel hybrid simulator inside the
filter therefore avoids the need to evaluate the transition density associated
with the hybrid model. We believe that this is the first serious attempt to
explore the performance of a hybrid simulator when used as an inferential tool.

To validate the methodology, we apply the method to an autoregulatory process
with five reactions and two species. This simple application allows comparison
of the proposed hybrid inference scheme with a scheme for performing inference
for the true underlying discrete stochastic model. Finally, we compare the
performance of the proposed hybrid scheme as an inferential tool with an
approach based upon the simulation methodology described in
\citeasnoun{salis2005}.

The remainder of the article is structured as follows. In Section~\ref{kin} we
give a brief exposition of the stochastic approach to chemical kinetics before
outlining the hybrid simulation technique in Section~\ref{sim}. 
Section~\ref{particle} describes the particle MCMC scheme for inference. This is
then applied in Section~\ref{app} before conclusions are drawn in
Section~\ref{disc}.

\section{Stochastic Kinetics -- A Brief Review}\label{kin}

We consider here the stochastic approach to chemical kinetics and
outline a Markov jump process (MJP) description of the dynamics of a system of
interest, expressed by a reaction network. Two approximations that can be used
in a hybrid modelling approach are outlined. For further details regarding
stochastic kinetics we refer the reader to \citeasnoun{wilkinson2012}.

\subsection{Stochastic Kinetic Models}

A biochemical network is represented with a set of reactions. We have $k$
species $\mathcal{X}_{1},\mathcal{X}_{2},\ldots ,\mathcal{X}_{k}$ and $r$
reactions $R_{1},R_{2},\ldots ,R_{r}$ with a typical reaction $R_{i}$ of the
form,
\[
\begin{array}{cccc}	
  R_{i}: \,  & u_{i1}\mathcal{X}_{1}+\ldots +u_{ik}\mathcal{X}_{k} &\xrightarrow{\phantom{a}c_{i}\phantom{a}} 
  & v_{i1}\mathcal{X}_{1}+\ldots +v_{ik}\mathcal{X}_{k}.	
\end{array}
\]
Note that $c_{i}$ is the kinetic rate constant associated with reaction $R_{i}$ and we write 
the vector of all rate constants as $\bmc=(c_{1},c_{2},\ldots ,c_{r})'$. Clearly, the effect 
of reaction $i$ on species $j$ is to change the number of molecules of $\mathcal{X}_{j}$ by an amount 
$v_{ij}-u_{ij}$. To this end, we may define the $r\times k$ \textit{net effect} matrix $\bmA$, given by 
$\bmA=\left\{a_{ij}\right\}$ where $a_{ij}=v_{ij}-u_{ij}$. To induce a compact notation, 
let $\bmX(t)=(X_{1}(t),X_{2}(t),\ldots ,X_{k}(t))'$ denote the number 
of molecules of each respective species at time $t$. Now, under the assumption of mass action kinetics, 
the instantaneous hazard of $R_{i}$ is 
\[
h_i(\bmX(t),c_i) = c_i\prod_{j=1}^k \binom{X_{j}(t)}{u_{ij}}.
\]
The \textit{order} of reaction $i$ is $\sum_j u_{ij}$. The evolution of a
biochemical network of interest is most naturally modelled as a Markov jump
process. Whilst the transition density associated with the process typically
does not permit analytic tractability, the process can be exactly simulated
forwards in time using a discrete event simulation method. The 
most well-used method is known
in the stochastic kinetics literature as the \textit{Gillespie algorithm}
\cite{gillespie1977} and uses the fact that if the current time and state are
$t$ and $\bmX(t)$ respectively then the time $\tau$ to the next reaction event
is
\[
\tau\sim \textrm{Exp}\left\{\lambda(\bmX(t),\bmc)\right\}, \quad \textrm{where \, $\lambda(\bmX(t),\bmc)=\sum_{i=1}^{r}h_{i}(\bmX(t),c_{i})$},
\]   
and the reaction that occurs will be type $R_{i}$ with probability proportional
to the reaction hazard $h_{i}(\bmX(t),c_{i})$. Other exact simulation methods
are possible -- Gibson and Bruck's next reaction method \cite{gibson2000} is
widely regarded to be the most computationally efficient strategy. As these
methods capture every reaction occurrence, they can be extremely computationally
costly for many systems of interest.

\subsection{Chemical Langevin Equation}

The CLE \cite{kampen2001,golightly2005} can be constructed by calculating 
the infinitesimal mean and variance of the Markov jump process and matching these quantities to the 
drift and diffusion coefficients of an It\^o stochastic differential equation (SDE). If we write
$d\bmX(t)$ for the $k$-vector giving the change in state of each species in the
time interval $(t,t+dt]$ then $d\bmX(t)=\bmA'd\bmR(t)$ where $d\bmR(t)$ is the
$r$-vector whose $i$th element is a Poisson random quantity with mean
$h_{i}(\bmX(t),c_{i})dt$. Hence, we arrive at
\[
E\left\{d\bmX(t)\right\}= \bmA'\bmh(\bmX(t),\bmc)dt,\qquad Var\left\{d\bmX(t)\right\}=\bmA'\textrm{diag}\left\{\bmh(\bmX(t),\bmc)dt\right\}\bmA,
\]
where $\bmh(\bmX(t),\bmc)=(h_{1}(\bmX(t),c_{1}),\ldots ,h_{r}(\bmX(t),c_{r}))'$
is the $r$-vector of hazards. Consequently, the It\^o SDE with the same infinitesimal mean and variance as the true Markov
jump process is
\begin{equation}\label{da}
  d\bmX(t)=\bmA'\bmh(\bmX(t),\bmc)\,dt + \sqrt{\bmA'\textrm{diag}\left\{\bmh(\bmX(t),\bmc)\right\}\bmA}\,d\bmW(t),
\end{equation}
where $d\bmW(t)$ is the increment of a $k$-dimensional Brownian motion and \linebreak[4]
$\sqrt{\bmA'\textrm{diag}\left\{\bmh(\bmX(t),\bmc)\right\}\bmA}$ is
any $k\times k$ matrix
square root. Note that ignoring the driving noise term in (\ref{da}) will yield
the deterministic ordinary differential equation (ODE) representation of the
system. The SDE in (\ref{da}) will be typically analytically intractable and it is therefore natural to work with 
the Euler-Maruyama approximation
\begin{equation}\label{euler}
  \Delta \bmX(t)=\bmA'\bmh(\bmX(t),\bmc)\,\Delta t+\sqrt{\bmA'\textrm{diag}\left\{\bmh(\bmX(t),\bmc)\right\}\bmA}\,\Delta \bmW(t)
\end{equation}
where $\Delta \bmW(t)\sim \textrm{N}(0,\bmI\Delta t)$. Given the intractability of the CLE, we eschew this approach in favour 
of a further approximation which generally processes a greater degree of tractability than the CLE. This linear noise approximation 
(LNA) is the subject of the next section.

\subsection{Linear Noise Approximation}

The LNA can be viewed either as an approximation to the 
MJP or CLE and consequently can be obtained in a number 
of more or less formal ways. Here, we derive the LNA as a general 
approximation to the solution of an arbitrary SDE before considering 
the specific SDE given by the CLE. For further details of the LNA, 
we refer the reader to \citeasnoun{Komorowski09} and \citeasnoun{sherlock2012} for recent discussions.  

Consider now the SDE satisfied by an It\^o process $\{\bmX(t)\}$ of length $k$, 
\begin{equation}\label{eqn.full.sde}
  d\bmX\oft = \bmalpha(\bmX\oft) \, dt + \epsilon\bmbeta(\bmX\oft)\, d\bmW\oft,
\end{equation}
with initial condition $\bmX\oftime{0}=\bmx_0$. Let $\bmeta\oft$ be the
(deterministic) solution to
\begin{equation}\label{eqn.deterministic.y}
  \frac{d\bmeta}{dt} = \bmalpha(\bmeta)
\end{equation}
with initial value $\bmeta_0$. We assume that over the time interval of interest
$\Norm{\bmX-\bmeta}$ is $O(\epsilon)$. Set
$\bmM\oft=(\bmX\oft-\bmeta\oft)/\epsilon$ and Taylor expand $\bmX\oft$ about
$\bmeta\oft$ in (\ref{eqn.full.sde}). Collecting terms of $O(\epsilon)$ gives
\begin{equation} \label{eqn.perturb}
d\bmM\oft = \bmF\oft \bmM\oft\,dt + \bmbeta\oft \,d\bmW\oft,
\end{equation}
where $\bmF$ is the $k \times k$ matrix with components
\[
F_{ij}\oft=\left.\frac{\partial \alpha_i}{\partial x_j}\right|_{\bmeta\oft}
\quad\text{and}\quad
\bmbeta\oft=\bmbeta(\bmeta\oft).
\]
The initial condition for (\ref{eqn.perturb}) is $\bmM(0)=(\bmx_0-\bmeta_0)$,
and thereafter $\bmM\oft$ is Gaussian for all $t$, provided that the initial
condition is a fixed point mass or follows a Gaussian distribution. The 
$\epsilon$ in (\ref{eqn.full.sde}) indicates that the intrinsic noise
term $\epsilon\bmbeta(\bmX\oft)$ is ``small'', but plays no part 
in the form of \eqref{eqn.perturb}. For simplicity of presentation, 
therefore, and without loss of generality we henceforth set $\epsilon=1$.

Suppose now that $\bmM\oftime{0} \sim \textrm{N}(\bmm_0,\bmV_0)$; in
this case the SDE satisfied by $\bmM(t)$ in equation (\ref{eqn.perturb}) can be solved
analytically (see Appendix \ref{lnaSol}) to give
\begin{equation} \label{lna.solution1}
\bmM\oft 
\sim 
\textrm{N} \left(\bmG\oft \bmm_0, \bmG\oft \bmPsi\oft \bmG\oft' \right).
\end{equation}
Here $\bmG$ is the fundamental matrix for the deterministic ODE
$d\bmm/dt = \bmF\oft\bmm$, so that
\begin{equation} \label{lna.solution2}
\frac{d\bmG}{dt} = \bmF\oft\bmG; \quad \bmG\oftime{0}=\bmI,
\end{equation}
and $\bmPsi$ satisfies
\begin{equation} \label{lna.solution3}
\frac{d\bmPsi}{dt} =
\bmG^{-1} \oft \bmbeta \oft \bmbeta\oft'\left(\bmG^{-1}\oft\right)';
\quad \bmPsi\oftime{0}=\bmV_0.
\end{equation}
Hence we obtain
\[
\bmX\oft\sim\textrm{N}\left(\bmeta\oft+\bmG\oft\bmm_0, \bmG\oft\bmPsi\oft\bmG\oft'\right).
\]
In the following, we aim to exploit the analytic tractability of the LNA to build a
novel hybrid model allowing both efficient simulation and inference.

\section{Hybrid Simulation via the LNA}\label{sim}

Hybrid simulation strategies begin by partitioning the reactions into two
subsets, ``fast'' and ``slow''. It is helpful at this point to also label any
\textit{species} that are changed by one or more fast reactions as fast and the
remaining species as slow. In between any two slow reaction events we model the 
dynamics of each species changed by the action of a fast reaction via the LNA. 
Since the slow reaction hazards will, in general, depend on
species changed by fast reaction occurrences, slow reaction event times will
follow an inhomogeneous Poisson process. We simulate slow reaction events via
thinning \cite{lewis1979}, which requires an upper bound on the total slow
reaction intensity.

In the following section, we give a novel dynamic re-partitioning scheme and
provide a justification of the approach. In Section~\ref{bound}, we derive a
probable bound on a linear combination of LNA components before using this
result to give a probable upper bound on the total intensity of all slow
reactions in Section~\ref{slowbound}. We describe our hybrid simulation strategy
algorithmically in Section~\ref{alg}.

\subsection{Choice of reaction type}\label{reactchoice}

Consider the general criterion that over some time interval $\Delta t$ the
changes brought about by reaction $j$ have a small relative impact on the state
vector, $\bmX$; such changes will also have a small relative impact on the rate
of each reaction. We represent a typical number of occurrences of a reaction by
its expectation; however even if this expectation is less than one, we do not
wish a single occurrence of $j$ to cause a substantial change in the state
vector. For a reaction $j$ to be regarded as fast, we therefore require
\begin{equation}\label{eqn.hybrid.choice.a}
\Abs{a_{ji}} \max\left(1,h_j\Delta t\right)\le \epsilon X_i\quad 
\end{equation}
for all $i$ such that $a_{ji}\neq0$ and for some $\epsilon >0$ which represents ``small''. 

Our proposed scheme re-evaluates the choice of reactions which can safely be
modelled as fast at intervals of at most $\Delta t_{hybrid}$. Clearly this
choice must be valid until the next re-evaluation and so, we require
\eqref{eqn.hybrid.choice.a} to hold with $\Delta
t_{hybrid}$ and $\epsilon$ equal to some $\epsilon_{hybrid}$.

Both the CLE and LNA are based upon the Gaussian approximation to the Poisson
distribution; let us deem this approximation to be sufficiently accurate
provided that the mean of the Poisson distribution is at least $N^*$. We
therefore require that, \textit{over the time interval where changes brought
  about by reaction $j$ start to noticeably affect the rates of at least one
  reaction} (which may be reaction $j$), \textit{the mean number of occurrences
  of reaction $j$ should be at least $N^*$}. Let $\Delta t_j$ be the time
interval over which changes brought about by reaction $j$ start
  to have an effect. Now for some suitable choice of $\epsilon=\epsilon^*$, $\Delta t_j$ is
the largest value $\Delta t$ which satisfies (\ref{eqn.hybrid.choice.a}).
Clearly if $\Abs{a_{ji}} > \epsilon^* X_i$ for at least one $i$ then
(\ref{eqn.hybrid.choice.a}) cannot be satisfied and the reaction must be slow.
Otherwise $\Delta t_j$ is the largest $\Delta t$ that satisfies $\Abs{a_{ji}}
h_j \Delta t\le \epsilon^* X_i~\forall i$; i.e. $h_j \Delta t_j = \epsilon^*
\min_i \frac{1}{\Abs{a_{ji}}}X_i$. We however need $h_j \Delta t_j \ge N^*$; for
an equation to be considered as fast we must therefore require that
\begin{equation}\label{eqn.hybrid.choice.b}
\Abs{a_{ji}}N^* \le \epsilon^* X_i 
\end{equation}
for all $i$ such that $a_{ji}\neq0$. As might be inferred from the italicised
fundamental condition, $\Delta t_j$ does not appear explicitly in this equation.
Note also that subject to (\ref{eqn.hybrid.choice.b}), the requirement in
(\ref{eqn.hybrid.choice.a}) $\Abs{a_{ji}} \le \epsilon X_i~\forall i$ is
automatically satisfied provided $\epsilon\ge\epsilon^*/N^*$.

In summary, for reaction $j$ to be classified as fast, we require
(\ref{eqn.hybrid.choice.b}) to be satisfied, and (\ref{eqn.hybrid.choice.a}) to
be satisfied for $\Delta t=\Delta t_{hybrid}$ and $\epsilon=\epsilon_{hybrid}$.

\subsection{Probable bounds on a linear combination of
  LNA components}\label{bound}

An upper bound on the total intensity of all slow reactions can be found by
deriving an upper bound on a linear combination of the components that drive the
LNA. We therefore require an upper bound of a function of the form $\sum_{i=1}^k
b^*_i\oft M_i\ofr$, $r \in[0,t]$, where $\bmM\ofr$ satisfies
(\ref{eqn.perturb}). The following result provides a bound which holds with
probability as close to $1$ as desired. A proof can be found in
\ref{boundproof}.

\begin{proposition}
  Let $M_i\oft$, $i=1,\ldots, k$ be the components of the stochastic vector
  $\bmM\oft$ which satisfies $\bmM\oftime{0}=\bmzero$ and evolves according to
  (\ref{eqn.perturb}). Define
\begin{equation}\label{eqn.tau}
\tau_i\oft:=\int_0^t\sum_{j=1}^{k}\left[\bmG^{-1}\ofr\bmbeta\ofr\right]^2_{ij}dr,
\end{equation}
where $\bmG(t)$ is the deterministic matrix defined in (\ref{lna.solution2}).
Set $\bmb\oft=\bmG\oft'\bmb^*\oft$, and
\begin{equation}\label{eqn.define.bstar}
b^{\max}_i:=\max_{r\in [0,t]}\Abs{b_i\ofr}, i=1,\ldots,k.
\end{equation}
For any $\epsilon \in (0,1)$ and every $i$ in $1,\ldots,k$ define
\begin{equation}\label{eqn.define.ustar}
u^*_i:=
-\Phi^{-1}\left(\frac{\epsilon}{4k}\right)\tau_i^{1/2},
\end{equation}
where $\Phi(\cdot)$ is the cumulative distribution function of a standard normal
distribution. Then
\[
\Prob{\max_{r\in[0,t]}\sum_{i=1}^k b^*_i\ofr M_i\ofr \le \sum_{i=1}^k
  b^{max}_iu_i^*} \ge 1-\epsilon.
\]

\end{proposition}

\subsection{Maximum intensity over an interval}\label{slowbound}

The evolution of species numbers that arises from fast reactions is modelled via
the LNA, whereas changes in species numbers that arise from slow reactions are
modelled though the Markov Jump process. In order to efficiently simulate slow
reaction events we require a relatively tight upper bound on the total hazard
(or intensity) of all slow reactions.

Consider the time interval between a given slow reaction event and either
the next slow reaction or the time ($\Delta t_{hybrid}$ in the future) when
reactions may be reclassified. Over this interval the number of molecules of
each slow species remains fixed, with changes in reaction hazards depending only
on the evolution of the relevant fast species. A first order reaction where the
rate depends only on the number of molecules of a single slow species may
therefore be treated, over this interval, as zeroth order, but with a different
rate constant. Similarly a second order reaction where one or both of the
reacting species are slow can be treated as a first or zeroth order reaction
over this interval. In common with most reaction models (e.g.
\citeasnoun{wilkinson2012}) we will assume that any apparent interactions
between more than two molecules are built up from reactions of order two or
fewer. For this interval we therefore partition the slow reactions into three
classes $R_s^{(0)}$, $R_s^{(1)}$ and $R_s^{(2)}$, for reactions which, over this
interval can be treated as zeroth, first and second order respectively, and
where these classifications are understood to depend on the current
classification of reactions into slow and fast.

Denoting by $X_k$ the number of molecules of species $k$, we therefore have
$h_j\left(t,c_j\right)=c^*_j$ for $j\in R_s^{(0)}$;
$h_j\left(t,c_j\right)=c^*_jX_{k_1(j)}$ for $j\in R_s^{(1)}$; and
$h_j\left(t,c_j\right)=c^*_jX_{k_1(j)}X_{k_2(j)}$ for $j\in R_s^{(2)}$,
%\[
%h_j\left(t,c_j\right)=c^*_j~\left(j\in R_s^{(0)}\right);~~
%h_j\left(t,c_j\right)=c^*_jX_{k_1(j)}~\left(j\in R_s^{(1)}\right);~~
%h_j\left(t,c_j\right)=c^*_jX_{k_1(j)}X_{k_2(j)}~\left(j\in R_s^{(2)}\right),
%\]
where $k_1(j)$ and $k_2(j)$ are the indices of the first and second (if
required) reactants involved in reaction  $j$, and each coefficient, $c_j^*$, is
proportional to the true rate constant, $c_j$, but also takes into account the
number of molecules of any slow reactants in reaction $j$.

Writing $X_i(t)=\eta_i(t)+M_i(t)$ and neglecting terms in $M_iM_j$, the total
intensity of all slow reactions is
\begin{align*}
  \lambda^{(s)}(\bmX(t))
  &\approx \sum_{j \in R_s^{(0)}} c^*_j+ \sum_{j \in R_s^{(1)}} c^*_j \left(\eta_{k_1(j)}(t) + M_{k_1(j)}(t)\right)\\
  &+ \sum_{j \in R_s^{(2)}} c^*_j \left(\eta_{k_1(j)}(t)\eta_{k_2(j)}(t) +
    \eta_{k_1(j)}(t)M_{k_2(j)}(t)+ \eta_{k_2(j)}(t)M_{k_1(j)}(t)\right)\\
  &= \lambda^{(s)}(\bmeta(t)) + \hspace{-.2cm}\sum_{j \in R_s^{(1)}} c^*_j M_{k_1(j)}(t) \\
  &\qquad +\sum_{j \in R_s^{(2)}} c^*_j \left( \eta_{k_1(j)}(t)M_{k_2(j)}(t) +
    \eta_{k_2(j)}(t)M_{k_1(j)}(t)
  \right).\\
\end{align*}
This can be rewritten as 
\begin{equation}\label{eqn.total.lambda}
\lambda^{(s)}(X(t))\approx\lambda^{(s)}(\bmeta(t))+\sum_{i=1}^kb^*_i\left(\bmc^*,\bmeta\oft\right)M_i\oft,
\end{equation}
where 
\begin{equation}\label{eqn.define.b}
b^*_i(\bmc^*,\bmeta\oft) =
\sum_{\{j \in  R_s^{(1)}:k_1(j)=i\}}\hspace{-.4cm}c^*_j+ 
\sum_{\{j\in R_s^{(2)}: k_2(j)=i\}}\hspace{-.4cm}c^*_j~\eta_{k_1(j)}
+
\sum_{ \{j\in R_s^{(2)}: k_1(j)=i\}}\hspace{-.4cm}c^*_j~\eta_{k_2(j)}.
\end{equation}
Note that the approximation in (\ref{eqn.total.lambda}) is exact if, over the
interval, all reactions can be treated as zeroth or first
order. Also $b_i=0$ if all reactions whose rate is influenced by 
species $i$ can be treated as zeroth order reactions over the time 
interval.

Defining $b^{max}_{i}$ and $u^*_{i}$ as in (\ref{eqn.define.bstar}) and
(\ref{eqn.define.ustar}) and, given that we choose to make $M_i\oftime{0}=0$, we
may therefore provide the following probable upper bound over the interval
$[0,T]$ on the total intensity of all slow reactions combined:
\begin{equation}\label{eqn.hmax}
  h^{s}_{\max}:=\lambda^{s}_{max} + \sum_{i=1}^{k}b_{i}^{max}u_{i}^{*},
\end{equation}
where
\[
\lambda^{s}_{max}:=\max_{t\in[0,T]}\lambda^{s}\left(\bmc^*,\bmeta\oft\right).
\]

\subsection{Generic Algorithm}\label{alg}

We now present a generic algorithm for simulating from a mixture of slow and
fast reactions using the Linear Noise Approximation for the fast reactions and
allowing the slow reactions to evolve through the ``exact'' Markov jump process.

Given a starting state the algorithm chooses a time interval, $\Delta
t_{integrate}$, over which to integrate the fast reaction mechanism and hence
detect whether or not there has been a potential slow reaction. If there is a
potential slow reaction in this interval then the fast reactions must be
reintegrated up to this potential slow reaction time to simulate the state
vector at this time. If the next slow reaction were to occur some considerable
time in the future then $[t_{curr},t_{curr}+\Delta t_{integrate}]$ would ideally
just fail to include this reaction time, and thereby eliminate the need to
re-integrate over such a large time interval. By contrast the penalty to
computational efficiency is smaller if there is just a small time interval until
the next potential slow reaction. However the upper bound on the total slow
intensity, and hence the rate at which potential reactions occur, increases with
$\Delta t_{integrate}$. Given the circularity of these constraints we simply set
$\Delta t_{integrate}$ as an arbitrary tuning factor. Furthermore, since we may
only re-evaluate the fast/slow status of each reaction at the end of an
integration we require $\Delta t_{integrate} \le \Delta t_{hybrid}$.

The algorithm commences at time $t_{curr}=0$ with an initial state vector of
$\bmx_{curr}:=(x_{curr,1},\dots,x_{curr,k})$ and ends at some pre-defined time
$t_{end}>0$ with $\bmx_{curr}$ corresponding to the the state vector at
$t_{end}$. The rate constants $\bmc$ are assumed to be known but to simplify our
presentation of the algorithm we remove explicit mention of $\bmc$ from the
notation. The algorithm starts with $\Delta t_{integrate}$ and $\Delta
t_{hybrid}$ set to their default (user-defined) values.
\begin{enumerate}
\item \label{step.start}\textbf{If} $t_{curr}\ge t_{end}$ then \textbf{stop}.
\item \label{step.nocheck} Set $\Delta t_{hybrid}=\min(\Delta
  t_{hybrid},t_{end}-t_{curr})$ and $\Delta t_{integrate}=\min(\Delta
  t_{integrate},t_{end}-t_{curr})$.
\item \label{step.classify} \textit{Classify reactions}: given $\bmx_{curr}$
  classify each reaction as either slow or fast.
\item \label{step.integrate} \textit{Preliminary integration over full
    interval}: integrate jointly over\linebreak $[t_{curr}, t_{curr}+\Delta
  t_{integrate}]$ the $k$-vector ODE for $\bmeta(t)$,
  (\ref{eqn.deterministic.y}), the $k\times k$ matrix ODE for $\bmG(t)$,
  (\ref{lna.solution2}), the ODEs for $\bmPsi(t)$, (\ref{lna.solution3}), and
  the integral for $\tau_i(t_{curr},\Delta t_{integrate})$ ($i=1,\dots,k$),
  (\ref{eqn.tau}). Initial conditions for the ODEs are $\bmeta(0)=\bmx_{curr}$,
  $\bmG(0)=\bmI$ and $\bmPsi(0)=\bmzero$. So that only fast reactions contribute
  to the evolution, for the purposes of this integration set the rate of each
  slow reaction to zero.
\item Keep running maxima over the course of the ODE integration in order to
  calculate $\lambda^{s}_{max}$ and $b^{max}_{i}$ over the interval
  $[t_{curr},t_{curr}+\Delta t_{integrate}]$.
\item Calculate $u^*_{i}~(i=1,\dots,k)$ from (\ref{eqn.define.ustar}).
\item Simulate the first event time $t_*$ from a Poisson process which starts at
  $t_{curr}$ and has intensity $h^{s}_{max}$ as given in (\ref{eqn.hmax}).
\item \textbf{If} $t_*>t_{curr}+\Delta t_{integrate}$ then there is \textit{no
    potential slow reaction} in $[t_{curr},t_{curr}+\Delta t_{integrate}]$; set
  $t_{curr}=t_{curr}+\Delta t_{integrate}$ and simulate the state vector at this
  new time, $\bmx_{t_{curr}}$; \textbf{go to Step \ref{step.start}}.
\item \textit{Second integration}: integrate the ODEs from Step
  \ref{step.integrate} (except (\ref{eqn.tau})) forward over the interval
  $[t_{curr},t_*)$, again with the rate of each slow reaction set to zero. This
  provides the distribution of of the species just before time $t_*$,
  $\bmX\left(t_*^-\right)$, given that no slow reactions occurred up until this
  time. Hence simulate $\bmx\oftime{t_*^-}$ and set $\bmx_{curr}\leftarrow
  \bmx\oftime{t_*^-}$.
\item Calculate the probability that a slow reaction actually occurs at
  $t_{*}$,
  $\lambda^{s}\left(\bmx_{curr}\right)/h^{s}_{max}$, and hence
  simulate whether or not a slow reaction occurs at $t_*$.
\item \textbf{If} \textit{no slow reaction} occurs then set
  $t_{curr}=t_*$ and \textbf{go to Step \ref{step.nocheck}}.
\item \textit{Update from slow reaction:} simulate which slow reaction occurs
  using the following probabilities for $j \in R_s$.
\[
\Prob{\mbox{slow reaction } j|\mbox{slow reaction}} =
\frac{h_j\left(\bmx_{curr}\right)}{\lambda^{s}(\bmx_{curr})};
\]
update $\bmx_{curr}$ according to the net effects vector for the chosen slow
reaction.
\item \label{step.housekeep} Set $t_{curr}=t_*$ and \textbf{go to Step \ref{step.nocheck}}.
\end{enumerate}

\section{Bayesian Inference}\label{particle}

We consider here the task of performing inference for the kinetic
rate constants $\bmc$ given noisy measurements on the system state $\bmX(t)$ at
discrete time points. We aim to embed the hybrid simulation method outlined in 
Section~\ref{sim} inside a recently proposed particle MCMC algorithm to 
obtain an efficient inference scheme. 

\subsection{A Particle MCMC approach}

Suppose that the process $\bmX(t)$ is not observed exactly, rather, we have
(without loss of generality) noisy measurements $\bmY_{0:T}=\{\bmY(t):t=0,\ldots
,T\}$ observed on a regular grid. We assume that the true underlying process
$\bmX(t)$ is linked to $\bmY(t)$ via the density $\pi(\bmy(t)|\bmx(t))$.
Moreover, we assume that the observations are conditionally independent given
the latent process.

Rather than perform inference for the exact Markov jump process, we work with
the hybrid model, and kinetic rate constants $\bmc$ governing this approximate
model. Let $\bmX_{(0,T]}=\{\bmX(t):t\in(0,T]\}$ denote the complete process path
on $(0,T]$ and denote the marginal density of $\bmX_{(0,T]}$, under the
structure of the hybrid model, by $\pi_{h}(\bmx_{(0,T]}|\bmx(0),\bmc)$, since it
depends on the starting value $\bmx(0)$ and the rate constants $\bmc$. Note that
this density can be sampled from by executing the algorithm described in
Section~\ref{sim}. Let $\pi(\bmx(0))$ and $\pi(\bmc)$ denote the respective
prior densities for $\bmX(0)$ and $\bmc$. Fully Bayesian inference may proceed
by sampling
\[
\pi\left(\bmc,\bmx_{[0,T]}|\bmy_{0:T}\right) \propto
\pi\left(\bmc\right)\pi\left(\bmx(0)\right)
\pi_{h}\left(\bmx_{(0,T]}|\bmx(0),\bmc\right)
\prod_{i=0}^{T}\pi\left(\bmx(i)|\bmy(i)\right)\,.
\]
In this work, interest lies in the marginal posterior density
\begin{align}
  \pi\left(\bmc |\bmy_{0:T}\right) &= \int \pi\left(\bmc,\bmx_{[0,T]}|\bmy_{0:T}\right)\,d\bmx_{[0,T]}\nonumber \\
  &\propto \pi(\bmc)\pi(\bmy_{0:T}|\bmc) \label{target}\,.
\end{align}
Inference is problematic due to the intractability of the marginal likelihood
$\pi(\bmy_{0:T}|\bmc)$. We generate samples (\ref{target}) by appealing to a
special case of the particle marginal Metropolis Hastings (PMMH) scheme
described in \citeasnoun{andrieu2010} and \citeasnoun{andrieu09}. In brief, we
propose a new $\bmc^{*}$ using a suitable proposal kernel $q(\bmc^{*}|\bmc)$ and
run a particle filter targeting $\pi(\bmx_{[0,T]}|\bmy_{0:T},\bmc^{*})$ to
obtain the filter's estimate of marginal likelihood, denoted
$\hat{\pi}(\bmy_{0:T}|\bmc^{*})$. At iteration $i$ the proposed $\bmc^*$ is
accepted with probability
\begin{equation}\label{aprob}
\min\left\{1,\frac{\hat{\pi}(\bmy_{0:T}|\bmc^{*}) \pi(\bmc^{*})}{\hat{\pi}(\bmy_{0:T}|\bmc^{(i-1)}) \pi(\bmc^{(i-1)})}
  \times \frac{q(\bmc^{(i-1)} | \bmc^{*})}{q(\bmc^{*} | \bmc^{(i-1)})} \right\}\,.
\end{equation} 
After initialising the rate constants
and at iteration $i=0$ with $\bmc^{(0)}$, the algorithm proceeds as follows for $i\geq 1$:
\begin{enumerate}
\item Draw $\bmc^{*}\sim q(\cdot|\bmc^{(i-1)})$.
\item Run a particle filter targeting $\pi(\bmx_{[0,T]}|\bmy_{0:T},\bmc^{*})$,
  and compute $\hat{\pi}(\bmy_{0:T}|\bmc^{*})$, the filter's estimate of
  marginal likelihood.
\item With probability (\ref{aprob}) accept a move to $\bmc^{*}$ otherwise put
  $\bmc^{(i)}=\bmc^{(i-1)}$.
\end{enumerate}
The scheme as presented can be seen as a pseudo-marginal Metropolis-Hastings
method \cite{beaumont03,andrieu09b}. In particular, provided that the estimator
of marginal likelihood is non-negative and unbiased (or has a
constant positive multiplicative bias that
does not depend on $\bmc$), it is straightforward to verify that the method
targets the marginal $\pi(\bmc |\bmy_{0:T})$. We let $\bmu$ denote all random
variables generated by the particle filter and write the estimate of marginal
likelihood as $\hat{\pi}(\bmy_{0:T}|\bmc)=\pi(\bmy_{0:T}|\bmc,\bmu)$. By
augmenting the state space of the Markov chain to include $\bmu$ the acceptance
ratio in (\ref{aprob}) can be rewritten as
\[
\frac{\pi(\bmy_{0:T}|\bmc^{*},\bmu^{*})\pi(\bmu^{*}|\bmc^{*}) \pi(\bmc^{*})}{\pi(\bmy_{0:T}|\bmc^{(i-1)},\bmu^{(i-1)})\pi(\bmu^{(i-1)}|\bmc^{(i-1)}) \pi(\bmc^{(i-1)})}
  \times \frac{q(\bmc^{(i-1)} | \bmc^{*})\pi(\bmu^{(i-1)}|\bmc^{(i-1)})}{q(\bmc^{*} | \bmc^{(i-1)})\pi(\bmu^{*}|\bmc^{*}) }
\]
and we see that the chain targets the joint density
\begin{equation}\label{target2}
\pi(\bmc,\bmu|\bmy_{0:T})\propto \pi(\bmy_{0:T}|\bmc,\bmu)\pi(\bmu |\bmc) \pi(\bmc)\,.
\end{equation}
Marginalising (\ref{target2}) over $\bmu$ gives $\pi(\bmc|\bmy_{0:T})$ as a
marginal density. We note that if interest lies in the joint posterior density
of $\bmc$ and the latent path, the above algorithm can be modified to target
$\pi\left(\bmc,\bmx_{[0,T]}|\bmy_{0:T}\right)$. Essentially, the ancestors of
each particle must be stored to allow sampling of the particle filter's
approximation to $\pi(\bmx_{[0,T]}|\bmy_{0:T},\bmc^*)$. We refer the reader to
\citeasnoun{andrieu2010} for further details.

Step 2 of the PMMH scheme requires implementation of a particle filter for the
successive generation of samples from $\pi(\bmx_{[0,j]}|\bmy_{0:j},\bmc^{*})$
for each $j=0,1,\ldots ,T$. Note that up to proportionality, and for $j>0$
\[
\pi(\bmx_{[0:j]}|\bmy_{0:j})\propto \pi(\bmy(j)|\bmx(j))\pi(\bmx_{[0:j-1]}|\bmy_{0:j-1})\pi_{h}(\bmx_{(j-1,j]}|\bmx(j-1))
\]
where we have dropped $\bmc^{*}$ from the notation. Now suppose that we have an
equally weighted sample of points (or \textit{particles}) of size $N$ from
$\pi(\bmx_{[0:j-1]}|\bmy_{0:j-1})$. Denote this sample by
$\big\{\bmx_{[0:j-1]}^{k},k=1,\ldots ,N\big\}$. The bootstrap particle filter of
\citeasnoun{gordon1993} generates an approximate sample from
$\pi(\bmx_{[0:j]}|\bmy_{0:j})$ with the following importance resampling
algorithm:
\begin{enumerate}
\item For $k=1,2,\ldots ,N$, draw $\bmx_{(j-1,j]}^{k}\sim \pi_{h}(\cdot
  |\bmx(j-1)^{k})$ using the hybrid simulator and construct the extended path,
  $\bmx_{[0,j]}^{k}=\left(\bmx_{[0,j-1]},\bmx_{(j-1,j]}\right)$.
\item Construct and normalise the weights,
  \[
  w^{(j)}_{k}= \pi(\bmy(j)|\bmx(j)^{k})\,,\quad
  \tilde{w}^{(j)}_{k}=\frac{w^{(j)}_{k}}{\sum_{l=1}^{N}w^{(j)}_{l}}\,,
  \]
  where $k=1,2,\ldots ,N$.
\item Resample $N$ times amongst the $\bmx_{[0,j]}^{k}$ using the normalised
  weights as probabilities.
\end{enumerate}
In the case $j=0$, $\pi(\bmx(0)|\bmy(0))$ can be sampled by replacing Step 1 in
the algorithm above with $N$ iid draws from the prior $\pi(\bmx(0))$. Hence,
after initialising the particle filter with a sample from the prior, the above
sequence of steps can be performed as each observation becomes available, with
the posterior sample at one time point used as the prior for the next. By using
the hybrid simulator to generate proposals inside the importance resampler,
evaluation of the associated likelihood is not required when calculating the
importance weights and the only term that needs to be evaluated is the tractable
density associated with the measurement error. This setup is flexible and can be
used with any forward simulator such as the Gillespie algorithm or chemical
Langevin equation.

After all data points have been assimilated, the filter's estimate of 
the marginal likelihood is
\begin{equation}\label{ml}
 \hat{\pi}(\bmy_{0:T}) = \hat{\pi}(\bmy(0)) \prod_{j=0}^{T-1}\hat{\pi}(\bmy(j+1)|\bmy_{0:j})=\prod_{j=0}^{T}\frac{1}{N}\sum_{k=1}^{N}w_{k}^{(j)}
\end{equation}
for which we obtain unbiasedness under mild conditions involving the
resampling scheme, satisfied by the bootstrap filter described above
\cite{delmoral04}. Note that for the special case of the PMMH algorithm used
here, when running the particle filter, we need only store the values of the
latent states at each observation time, and each unnormalised weight.

\subsubsection{Tuning}\label{tuning}

The PMMH scheme requires specification of a number of particles 
$N$ to be used in the particle filter at Step 2. As noted by 
\cite{andrieu09b}, the mixing efficiency of the PMMH 
scheme decreases as the variance of the estimated marginal likelihood 
increases. This problem can be alleviated at the expense of greater 
computational cost by increasing $N$. This therefore suggests an optimal 
value of $N$ and finding this choice is the subject of \citeasnoun{pitt12}, \citeasnoun{doucet13} 
and \citeasnoun{sherlock2013}. The latter suggest that $N$ should be chosen so that the 
variance in the noise in the estimated log-posterior is around 2. \citeasnoun{pitt12} note that 
the penalty is small for a value between 0.25 and 2.25. We therefore recommend 
performing an initial pilot run of daPMMH to obtain an estimate of the 
posterior mean for the parameters $\bmc$, denoted $\hat{\bmc}$. The value 
of $N$ should then be chosen so that $\textrm{Var}(\log \pi(\bmy_{0:T}|\hat{\bmc}))$ 
is around 2.

In our application, we note that the rate constants $\bmc$ must be strictly
positive and we update $\log(\bmc)=(\log(c_{1}),\ldots ,\log(c_{r}))'$ in a single
block using a random walk proposal with Gaussian innovations. The innovation
variance must be chosen appropriately to maximise statistical efficiency through
well mixing chains. We take the innovation variance to be $\gamma
\hat{\textrm{var}}(\bmc)$, where $\hat{\textrm{var}}(\bmc)$ is obtained from a
short pilot run of the scheme. Following \citeasnoun{sherlock2013} we tune the
scaling parameter $\gamma$ to give an acceptance rate of approximately $10\%$.

\section{Application: Autoregulatory Network}\label{app}

To assess the performance of the proposed hybrid approach as a simulator and as
an inferential model, we consider a simple autoregulatory network with two
species, $\mathcal{X}_{1}$ and $\mathcal{X}_{2}$ whose time course behaviour
evolves according to the following set of coupled reactions,
\begin{align*}
  R_{1}:\quad  \emptyset  &\xrightarrow{\phantom{a}c_{1}\phantom{a}} \mathcal{X}_{1}  &
  R_{2}:\quad  \emptyset  &\xrightarrow{\phantom{a}c_{2}\phantom{a}} \mathcal{X}_{2}  \\
  R_{3}:\quad  \mathcal{X}_{1}  &\xrightarrow{\phantom{a}c_{3}\phantom{a}} \emptyset &
  R_{4}:\quad  \mathcal{X}_{2}  &\xrightarrow{\phantom{a}c_{4}\phantom{a}} \emptyset \\
  R_{5}:\quad  \mathcal{X}_{1}+\mathcal{X}_{2}  &\xrightarrow{\phantom{a}c_{5}\phantom{a}} 2\mathcal{X}_{2} 
\end{align*}
Essentially, reactions $R_{1}$ and $R_{2}$ represent immigration, reactions $R_{3}$ and
$R_{4}$ represent death and finally $R_{5}$ can be thought of as
interaction between the two species. Note that even for this simple system, 
the transition density associated with the resulting Markov jump process 
(under an assumption of mass action kinetics) cannot be found in closed form.

Throughout this section we take
\begin{equation}
\label{eqn.rate.formula}
\bmc=(2, sc, 1/50, 1, 1/(50\times sc))',
\end{equation}
and investigate the performance of our hybrid algorithm (henceforth designated as \textit{Hybrid LNA}) with regard to both
the simulated distribution of ${X}_1$ and ${X}_2$ and
inference on $\bmc$ for $sc\in\{1,10,100,1000\}$. The `probable upper
bound' of Section \ref{bound} is fixed to hold with probability
$1-10^{-6}$, whilst the relative and absolute errors of the stiff ODE solver
were set to $10^{-4}$.

We use the 
dynamic repartitioning procedure described in Section~\ref{reactchoice} with
$N^{*}=15$ and $\epsilon^{*}=\epsilon=0.25$. 
Reactions are reclassified as fast
or slow every $\Delta t_{hybrid}=\Delta t_{integrate}=0.1$ time units. For this
specification, Equation~(\ref{eqn.hybrid.choice.b}) ensures that a reaction will
be regarded as slow if the species numbers of species affected by that reaction are $60$ or
fewer.
The rates in \eqref{eqn.rate.formula} lead to an equilibrium for the MRE of
\[
[{X}_1,~{X}_2]=[50(1+sc-\sqrt{1+sc^2}),~1+\sqrt{1+sc^2}],
\]
which, for $sc \gg 1$ is approximately $[50-25/sc,~sc]$. Thus, for
$sc\gg 1$, when the system is at equilibrium, 
${X}_1$ is typically small, ${X}_2$ is
typically large, and reactions $R_2$ and $R_4$ are typically
fast. 

If $R_2$ and $R_4$ were always the only fast reactions and
$\mathcal{X}_2$ were always the only fast species then the LNA for the evolution of
${X}_2$ conditional on no slow reactions taking place would be
analytically tractable and, further, there would be no need for
dynamic repartitioning. We, however,
do not take advantage of this special case as we wish to show the generic
applicability of our method. To this end we also start each system
away from equilibrium, at $\bmX(0)=(0,0)'$.

For comparison, we also ran the Gillespie algorithm and a discrete/SDE 
hybrid simulation method in the spirit of the \textit{next reaction hybrid algorithm} 
of \citeasnoun{salis2005} (henceforth designated as \textit{Hybrid SDE}). Full 
details of this approach can be found in Appendix~\ref{hybsde}. 
For \textit{Hybrid SDE} we used the same dynamic partitioning criteria 
and additionally specified the required Euler time step to be 
$\Delta t_{Euler}=0.005$, which gave an accuracy comparable with 
that of \textit{Hybrid LNA}.

\subsection{Simulation}\label{comp.sim}

Using the autoregulatory network as a test case, we ran each hybrid
simulator  and the Gillespie
algorithm for $20,000$ iterations. 

\begin{figure}[t]
\centering
\includegraphics[width=\textwidth]{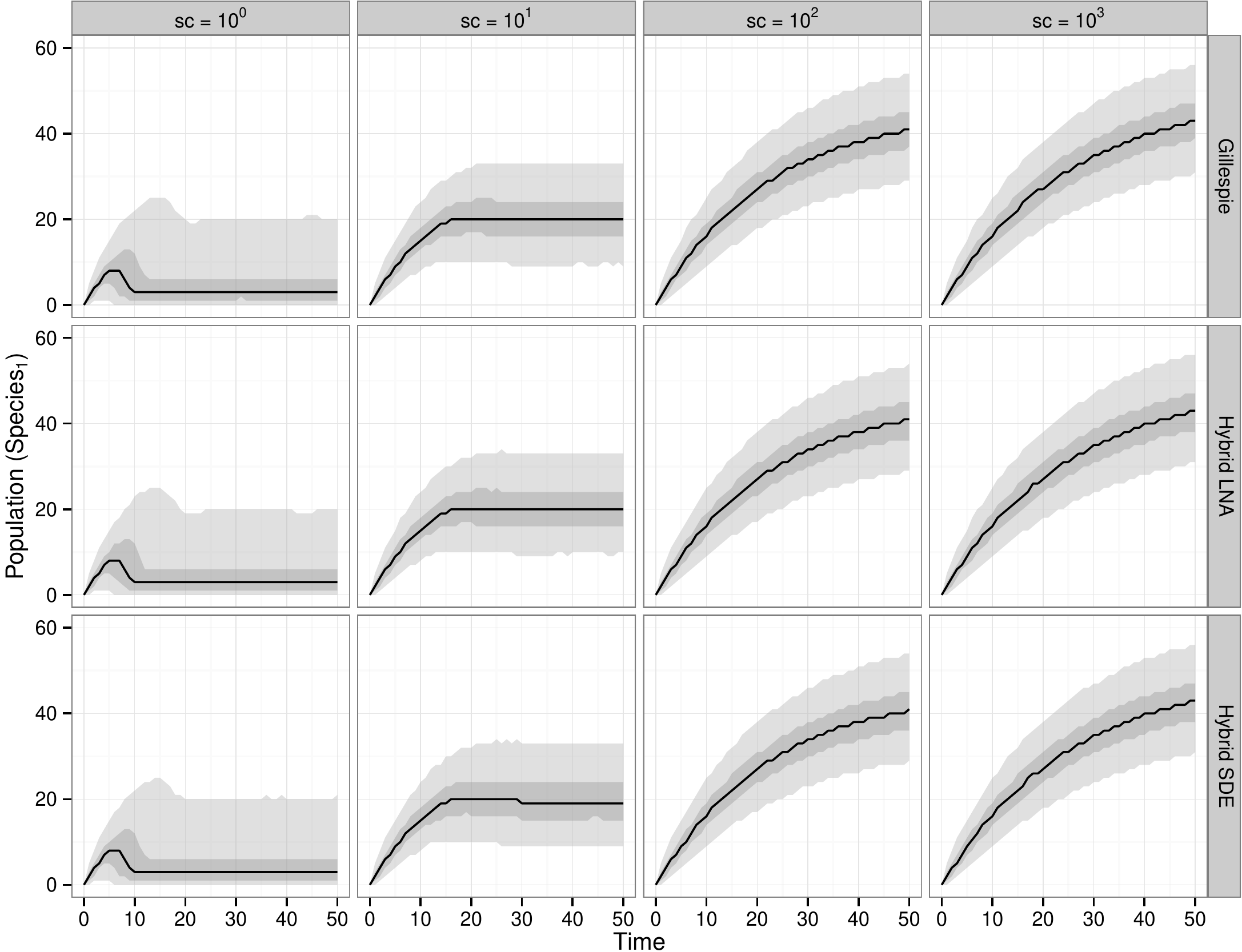}
\caption{Median (solid), inter-quartile range (inner shaded region) and 95\%
  credible region (outer shaded region) of $X_{1,t}$ based on $20,000$
  stochastic realisations of the model using Gillespie's direct method, Hybrid
  LNA and the Hybrid SDE. Model parameters were $(2, sc, 1/50, 1, 1/(50\times
  sc))'$.}\label{fig:sims}
\end{figure}

\begin{figure}[t]
  \centering
  \includegraphics[width=0.5\textwidth]{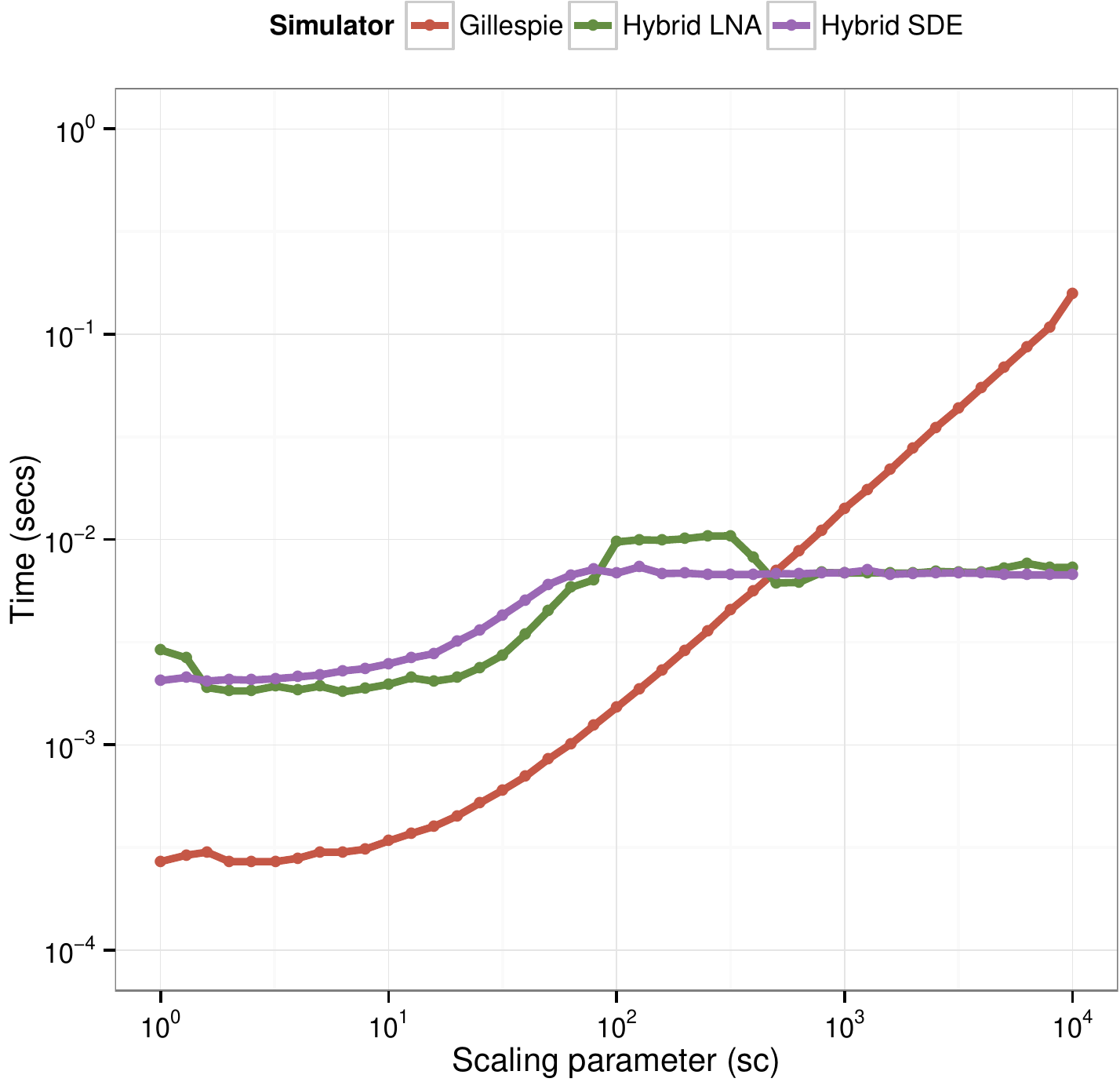}
  \caption{Simulator CPU time. Each point is the simulation time (in secs) of a
    single stochastic simulation, averaged over $1000$ simulations. Model
    parameters were $(2, sc, 1/50, 1, 1/(50\times sc))'$.}\label{fig:CPU}
\end{figure}

Figure~\ref{fig:sims} summarises the output of each simulation procedure, for
species $\mathcal{X}_{1}$ and Figure~\ref{fig:CPU} shows the CPU time of each
simulator, averaged over 1000 realisations (and using a much larger
set of values for $sc$. We see little difference between
simulator output. However, when taking into account computational cost, the
advantage of either hybrid approach over the Gillespie algorithm is clear. For
$sc<500$, reaction events occur relatively infrequently and the computational
cost of the hybrid algorithms is dominated by the computational overhead of
dynamic repartitioning. However for $sc>500$, the cost of both hybrid schemes is
roughly constant, whereas the cost of the Gillespie algorithm increases linearly
with $sc$. \textit{Hybrid LNA} requires minimal tuning, since the LNA
solution involves solving a set of ODEs, for which stiff solvers that
automatically and adaptively choose the time step 
so as to maintain a given level of accuracy  are
readily available. \textit{Hybrid SDE}, however, requires the 
user to choose a fixed Euler time-step, $\Delta t_{Euler}$, and manually 
attempt to balance accuracy against computational effort; moreover, since the
CLE is stiff and non-deterministic, there is the possibility that any fixed $\Delta
t_{Euler}$ might not maintain a desired level of accuracy throughout
repeated simulations, especially with different rate constants,
$\bmc$. Furthermore, the slow reaction updating procedure of Hybrid SDE can be
inefficient in a number of ways. The algorithm requires that only one slow
reaction event occurs in the interval over which the fast species are
integrated. If more than one slow reaction is detected, $\Delta t_{hybrid}$ is
reduced, the system state is rewound and a reclassification of reactions takes
place. Because of the reduction in $\Delta t_{hybrid}$, the system rewind may reclassify some erstwhile fast reactions as slow
and so actually increase the chance of multiple slow reaction occurrences.
Moreover, there is a subtle error in the algorithm: if a rewind 
has occurred, the new forward simulation must be conditional on the 
previously-simulated values of the fast reactants over the old 
interval of length $\Delta t_{hybrid}$. Strictly speaking therefore, 
these values should be stored and re-used, with approximate bridges 
constructed if it is necessary to fill in between the stored values. However 
if some of the previously-fast reactants have now become slow then it is not at all 
clear how to condition on the results from the previous attempt at 
forwards simulation. We therefore did not make make any attempt to 
correct this problem.

\subsection{Inference}\label{inf}

\begin{figure}[t]
  \centering
  \includegraphics[width=\textwidth]{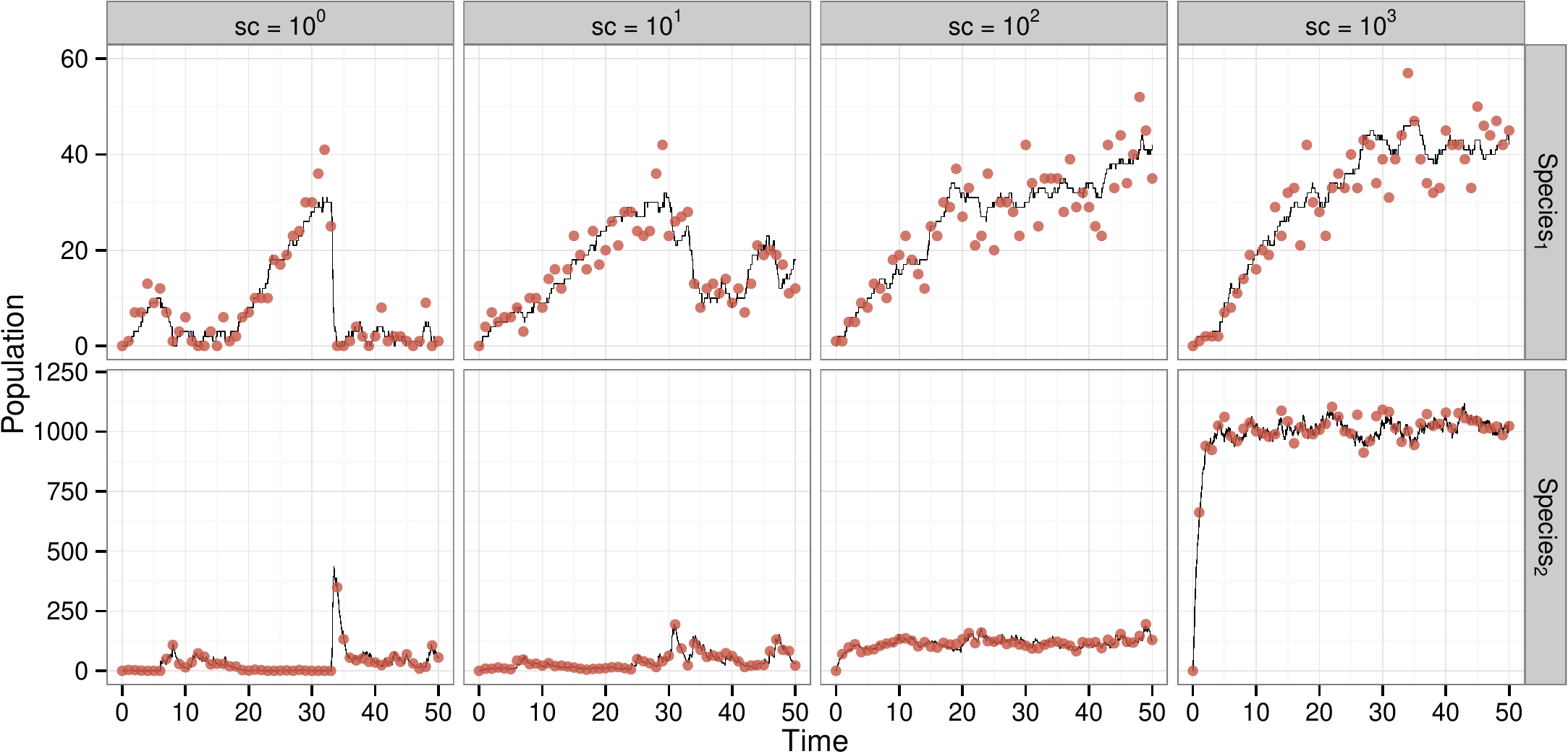}
  \caption{The four synthetic datasets used. Each data set was generated via the
      Gillespie algorithm. The true species
      numbers are represented by a black line. The noised observations
      are indicated by dots.}\label{fig:data}
\end{figure}

Data were simulated at integer times on $[0,50]$ via the Gillespie algorithm. This gave four
synthetic datasets which were then corrupted to give
  observations with a conditional distribution of
\[
Y_{i}(t)|X_{i}(t)\sim
\begin{cases}
  \textrm{Poisson}\left(X_{i}(t)\right) & \text{if $X_{i}(t)>0$},\\
  \textrm{Bernouilli}(0.1) & \text{if $X_{i}(t)=0$}
\end{cases}
\]
for each component $i=1,2$. The data 
are plotted in Figure~\ref{fig:data}, wherein, and for the remainder
of this section, we refer to the PMMH scheme that uses a given
  simulator by using the name of that simulator: \textit{Hybrid LNA},
  \textit{Hybrid SDE} and \textit{Gillepsie}.

To ensure identifiability, $c_{3}$ was fixed at
its true value, while independent Uniform $U(-8,8)$
priors were used for the remaining
$\log(c_i)$. For each combination of synthetic dataset and
  scheme  we performed a pilot run with 50 particles 
to obtain an approximate covariance matrix
$\hat{\textrm{Var}}(\bmc)$ and approximate posterior mean $\hat{\bmc}$.
Following the practical advice of \citeasnoun{sherlock2013}, further pilot runs
were performed with $\bmc$ fixed at $\hat{\bmc}$ to determine the number of
particles $N$ that gave a variance of the estimator of log-posterior $\log
\pi(\bmy_{0:T}|\hat{\bmc})$ of around $2$. Table~\ref{tab:tabpart} shows the
number of particles used for each scheme and each dataset. Note that \textit{Hybrid SDE}
required more particles than \textit{Hybrid LNA} or
\textit{Gillespie}, with nearly an order of magnitude difference
when $sc=1$. We found that using fewer particles would result in particle
degeneracy around time point 32, with only a few particles able to capture the
increase in $R_{5}$ occurrences around this time point.
  
\begin{table}
\centering
\begin{tabular}{@{} l lll @{}}
\toprule
& \multicolumn{3}{c}{Simulator} \\
\cmidrule(l){2-4}
sc & Gillespie & Hybrid$_{\text{{\tiny LNA}}}$ & Hybrid$_{\text{{\tiny SDE}}}$ \\
\midrule
$10^0$ & 250 & 250 & 1750 \\
$10^1$ & 800 & 800 & 1500 \\
$10^2$ & $\phantom{0}65$ & $\phantom{0}65$ & $\phantom{0}125$ \\
$10^3$ & $\phantom{0}65$ & $\phantom{0}65$ & $\phantom{00}85$ \\
\bottomrule
\end{tabular}
\caption{Number of particles used for each scheme and each synthetic dataset.}\label{tab:tabpart}
\end{table}

\begin{figure}[!t]
  \centering
  \includegraphics[width=\textwidth]{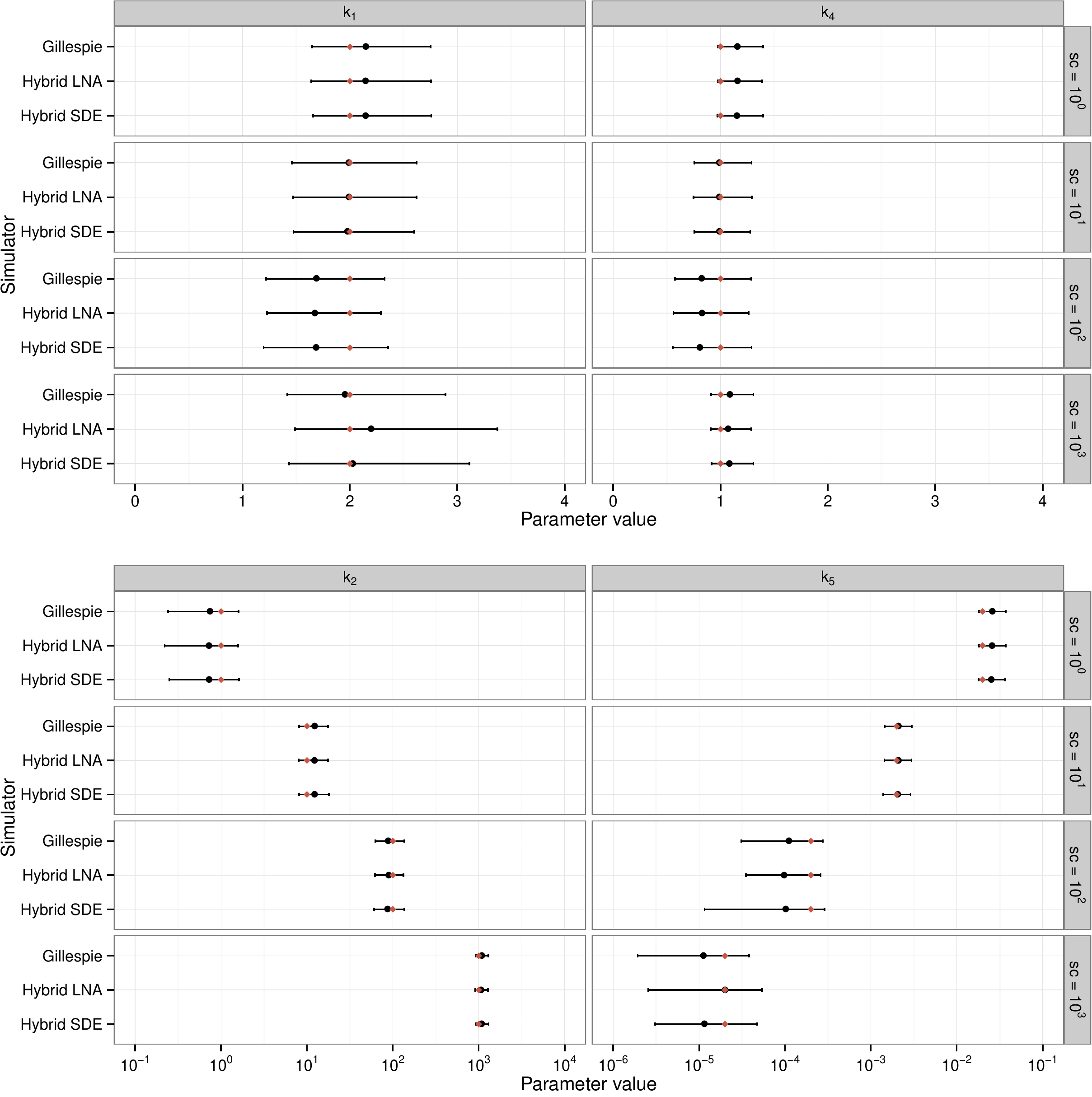}
  \caption{95\% credible regions and posterior medians (black dot) for each
    parameter value based on the output of each PMCMC scheme (Gillespie, Hybrid
    LNA and Hybrid SDE). True values are indicated by a red
    dot.}\label{fig:fig_hsim}
\end{figure}

\begin{figure}[]
  \centering
  \includegraphics[width=\textwidth]{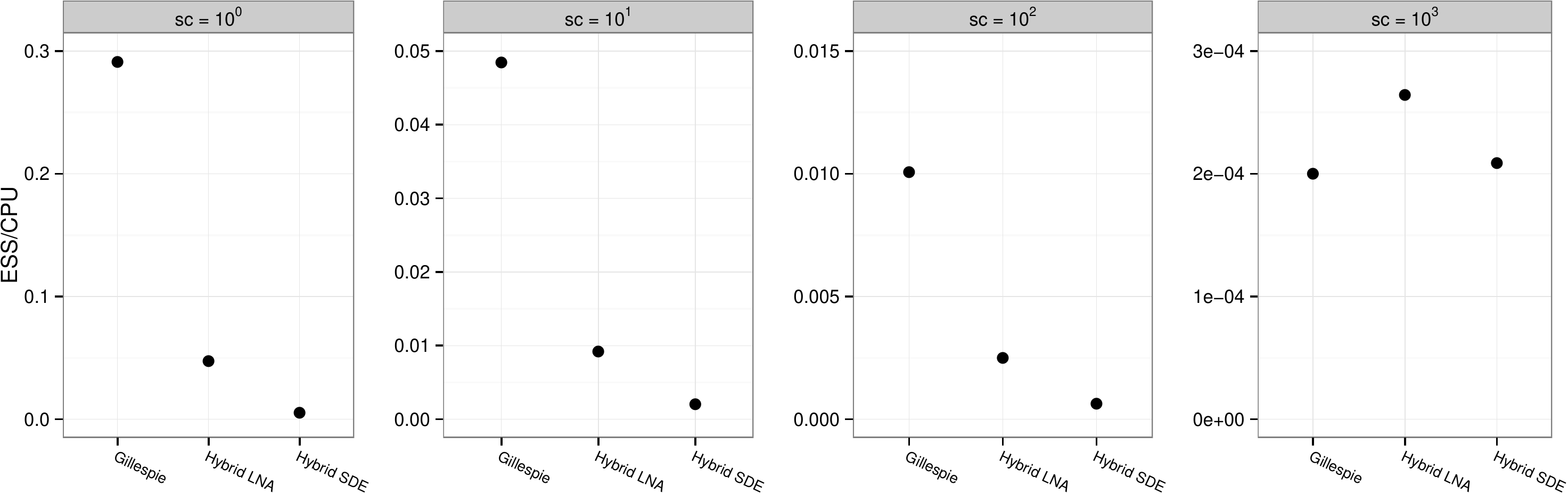}
  \caption{Minimum effective sample size (ESS) per second.}\label{fig:ess}
\end{figure}

We performed $2\times 10^{5}$ iterations of each scheme for $sc=1, 10, 100$ and
$2\times 10^{6}$ iterations for $sc=1000$. In all cases, the $\log(c_i)$ were
updated in a single block using a Gaussian random walk proposal kernel with an
innovation variance matrix given by $\gamma \hat{\textrm{Var}}(\bmc)$, with
$\gamma$ tuned to give an acceptance rate of around $10\%$.
Figure~\ref{fig:fig_hsim} summarises the posterior output of each scheme. We see
that in general, the sampled parameter values are consistent with the true
values that produced the data. There appears to be little difference between the
output of the PMMH scheme when using the Gillespie simulator, and both hybrid
schemes, suggesting that little is lost by adopting a hybrid model to perform
inference for the autoregulatory network. Figure~\ref{fig:ess} shows minimum
effective sample size (ESS) per second for each scheme. The results are
consistent with the timings shown in Figure~\ref{fig:CPU}. For relatively small
values of $sc$, reaction events occur relatively infrequently and little is to
be gained by running Hybrid SDE or Hybrid LNA over Gillespie. When using
$sc=1000$ we see a gain in overall efficiency for the hybrid schemes. We would
expect this relative gain to increase with $sc$, however, we found that the
computational cost of running the PMMH scheme with the Gillespie simulator
precluded comparison under this scenario.

\section{Discussion}\label{disc}

We have proposed a novel hybrid simulation method for efficiently simulating
stochastic kinetic models (SKMs). Our approach models fast reaction dynamics with the
LNA and slow dynamics with a Markov jump process. By deriving a probable upper
bound for a combination of components that drive the LNA, we obtain a probable
upper bound for the total slow reaction hazard thus allowing exact simulation of
the slow reaction events. This exactness is conditional on the accuracy of the upper bound,
of the LNA approximation and of the ODE solver
used to integrate the LNA. The first and the last of these were set
to high values, whilst the LNA itself is expected to be accurate
since it is only applied to reactions that are classified as fast. To
this end, reliable criteria for the
(dynamic) partitioning of reactions were also provided. Unlike existing
approaches to hybrid simulation that use the CLE, we avoid the need for a system
rewind (and the consequent difficulty in making the algorithm
  strictly correct). We also avoid the requirement to specify a
  fixed Euler time step which is unlikely to be appropriate across all
possible sets of rate parameters with prior support and all possible
realisations of the process.

We have also considered the task of inferring the rate constants governing SKMs
by adopting the hybrid model and performing exact simulation-based Bayesian
inference. We employed a recently-proposed particle MCMC scheme that, in its
simplest implementation, only requires the ability to forward simulate from the
model and evaluate an observation (or measurement error) density. We used this
scheme to compare results based on our proposed hybrid simulator with those obtained
under a hybrid simulator in the spirit of the work by \citeasnoun{salis2005}, and
also with inferences obtained under the ``exact'' Markov jump process
representation of the SKM. Both hybrid schemes led to inferences 
that were almost indistinguishable from those under the true model, 
with a clear indication of increasing relative efficiency as 
reaction rates increased.

\subsection*{Computing details}

All simulations were performed on a machine with 8GB of RAM and with an Intel i7
CPU. The operating system used was Ubuntu 12.04. The simulation code was mainly
written in C and compiled with flags: \texttt{-Wall}, \texttt{-O3},
\texttt{-DHAVE\_INLINE} and
\texttt{-DGSL\_RANGE\_CHECK\_OFF}. FORTAN code for the stiff ODE
  solver came from the \texttt{lsoda} package \cite{petzold83}. Graphics were constructed using R and the
ggplot2 R package \cite{R,ggplot2}.

The code can be downloaded from
\begin{center}
https://github.com/csgillespie/hybrid-pmcmc
\end{center}

\newpage

\appendix

\section{Appendices}

\subsection{Solution to the LNA}\label{lnaSol}

Recall that $\bmG$ is the fundamental matrix for the deterministic ODE
$d\bmm/dt=\bmF\oft\bmm$, satisfying equation~(\ref{lna.solution2}). Note that
\[
\bmzero=\frac{d}{dt}\bmG\bmG^{-1}=\bmG \frac{d\bmG^{-1}}{dt}
+\frac{d\bmG}{dt}\bmG^{-1},
~~\text{so}~\frac{d\bmG^{-1}}{dt}=-\bmG^{-1}\bmF\oft.
\]
Set
\[
\bmU\oft:=\bmG^{-1}\oft\bmM\oft,~~ \text{so}~ \bmU\oftime{0}=\bmM\oftime{0}.
\] 
Since $\bmG$ is deterministic, $d\bmG^{-1}d\bmM=\bmzero$ and so by
(\ref{eqn.perturb})
\[
d\bmU\oft=\bmG^{-1}\bmF\bmM
dt+\bmG^{-1}\bmbeta~d\bmW_t-\bmG^{-1}\bmF\bmM dt=\bmG^{-1}\bmbeta~d\bmW\oft.
\]
Thus
\[
\bmU\oft-\bmU\oftime{0}
=
\int_0^t\bmG^{-1}\ofr\bmbeta\ofr ~d\bmW\ofr.
\]
Therefore by linearity and Ito's Isometry,
\begin{equation}\label{eqn.Udist}
\bmU(t)-\bmU(0)\sim N\left(\bmzero,~\int_0^t\bmG^{-1}\ofr
\bmbeta\ofr\bmbeta\ofr'\left(\bmG^{-1}\ofr\right)' ~dr\right).
\end{equation}
Suppose now that $\bmM\oftime{0}~(=\bmU\oftime{0})\sim \textrm{N}(\bmm_0,\bmV_0)$, then
\begin{align*}
\bmM\oft 
&\sim
N
\left(
\bmG\oft \bmm_0, 
\bmG\oft
\bmPsi\oft
\bmG\oft'
\right)\\
\text{where}\quad
\bmPsi\oft
&=
\bmV_0+\int_0^t\bmG^{-1}\ofr\bmbeta\ofr\bmbeta\ofr'\left(\bmG^{-1}\ofr\right)'dr.
\end{align*}

\subsection{Proof of Proposition 1}\label{boundproof}

Firstly, $\sum_{i=1}^{k}b^*_i\ofr M_i\ofr = \sum_{i=1}^{k}b_i\ofr U_i\ofr$,
where $U_i$ is the $i^{th}$ component of the vector $\bmU$ defined in Appendix
\ref{lnaSol}, but with $\bmU(0)=\bmzero$ (since $\bmM(0)=\bmzero$). From its
definition, (\ref{eqn.tau}), $\tau_i$ is the $i^{th}$ diagonal component of the
variance in (\ref{eqn.Udist}), so
\[
\Prob{U_i(t)\ge u_i^*} = \Phi\left(-u_i^*/\sqrt{\tau_i}\right),
\]
for currently arbitrary values $u^*_i>0~,~~i\in \{1,\dots,k\}$.

Next, define the first hitting time $T_i(u^*_i)=\inf\{t:U_i(t)\ge u^*_i\}$. Now
$U_i(t)\ge u^*_i \Leftrightarrow T_i(u_i^*)\le t=0$ so
\[
\Prob{U_i(t)\ge u_i^*}=\Prob{U_i(t)\ge u_i^*|T_i(u_i^*)\le t}\Prob{T_i(u_i^*)\le t}.
\]
By the almost sure continuity of $U_i$, $\Prob{U_i(T_i(u_i^*))=u_i^*}=1$ and so
by the symmetry of $U_i$, $\Prob{U_i(t)\ge u_i^*|T_i(u_i^*)\le t}=1/2$. However
$T_i(u_i^*)\le t \Leftrightarrow \max_{(0,t]}U_i \ge u_i^*$, so
\[
\Prob{\max_{(0,t]}U_i \ge u_i^*}=2\Phi\left(-u_i^*/\sqrt{\tau_i}\right).
\]
Given some $\epsilon>0$, we may therefore choose
$u^*_i=-\Phi^{-1}\left({\epsilon}/{4k}\right)\tau_i^{1/2}$, which gives,
marginally,
\[
\Prob{\max_{(0,t]}U_i \ge u_i^*}=\frac{\epsilon}{2k}.
\]
By symmetry and the inclusion exclusion formula, therefore, marginally, 
\[
\Prob{\max_{(0,t]}\Abs{U_i} \ge u_i^*}=\frac{\epsilon}{k}.
\]
Hence
\[
\Prob{\Abs{U_i(r)}\le u_i^*:i\in\{1,\dots,k\},r\in(0,t]}=
1-\Prob{\max_{(0,t]}\Abs{U_i} \ge u_i^*~\text{for any}~i}\ge 1-\epsilon.
\]
Thus with probability at least $1-\epsilon$, for all $r \in [0,t]$
\[
\sum_{i=1}^kb^*_i\ofr M_i\ofr = \sum_{i=1}^kb_i\ofr U_i\ofr
\le
\sum_{i=1}^k\Abs{b_i\ofr}u_i^*
\le
\sum_{i=1}^kb^{max}_iu_i^*.
\]

\subsection{Hybrid Simulation based on the CLE}\label{hybsde}

We consider a hybrid simulation algorithm in the spirit of the \textit{next reaction hybrid 
algorithm} of 
\citeasnoun{salis2005}. This approach treats the subset of fast species with the
chemical Langevin equation and simulates their dynamics by numerically
integrating the corresponding SDE. Let $\bmX^{f}(t)$ be the state of the fast
species at time $t$. Suppose that we have $r^{f}$ fast reactions and $r^s$ slow
reactions. We then arrive at
\begin{equation}\label{daf}
  d\bmX^{f}(t)=\bmA_{f}'\bmh\big(\bmX(t),\bmc\big)\,dt+\sqrt{\bmA_{f}'\textrm{diag}
    \left\{\bmh^{f}\big(\bmX(t),\bmc\big)\right\}\bmA_{f}}\,d\bmW(t)
\end{equation}
where $\bmA_{f}$ is the $r^{f}\times k^{f}$ net effect matrix associated with
the fast reactions and $\bmh^{f}\big(\bmX(t),\bmc\big)$ is the $r^{f}$-vector of
fast reaction hazards which may depend on both fast and slow species numbers.
Hence, the fast specie numbers can be simulated by recursively iterating the
Euler discretisation of (\ref{daf}).

It remains that we can sample the times of the slow reactions. This step can be
performed by Monte Carlo, equating the integral of the time dependent
probability density for the time of the $j$th slow reaction to a uniform random
number. Since the slow reaction hazards are time varying, we write them as
$h_{j}^{s}(t,\bmc)$, $j=1,\ldots ,r^{s}$. Let $p_{j}(\tau_{j};t_{0})$ denote the
next reaction probability density for the $j$th slow reaction. Here, $t_{0}$ is
the time that the last occurred and $\tau_{j}$ is the time of the $j$th slow
reaction. From \citeasnoun{gibson2000}, $p_{j}(\tau_{j};t_{0})$ is a time
dependent exponential density
%\[
%p_{j}(\tau_{j};t_{0})=h_{j}^{s}(t_{0}+\tau_{j},c)\exp\left(-\int_{t_{0}}^{t_{0}+\tau_{j}}
%  h_{j}^{s}(t',c)dt'\right)
%\] 
for which the cumulative density function is
\begin{equation}\label{cdf}
  F(\tau_{j};t_{0})=1-\exp\left(-\int_{t_{0}}^{t_{0}+\tau_{j}}h_{j}^{s}(t',\bmc)dt'\right).
\end{equation}
Hence, setting equation (\ref{cdf}) equal to a uniform random number $r_{j}$ on
$(0,1)$ and simplifying gives
\begin{equation}\label{jump}
  \int_{t_{0}}^{t_{0}+\tau_{j}}h_{j}^{s}(t',\bmc)dt'+\log(r_{j})=0.
\end{equation}
We solve equation (\ref{jump}) by rearranging it in terms of a residual $R_{j}(t)$ and setting 
the integral upper bound to be a variable so that
\begin{equation}\label{jump2}
  \int_{t_{0}}^{t_{0}+t}h_{j}^{s}(t',\bmc)dt'+\log(r_{j})=R_{j}(t).
\end{equation}
Plainly, if $R_{j}(t)=0$ then $t=\tau_{j}$, $R_{j}(t)<0$ implies that $t<\tau_{j}$ and similarly 
if $R_{j}(t)>0$ then $t>\tau_{j}$. Hence, starting with state $\bmX(t)$ at time $t$, we can compute 
$\bmX(t+\Delta t)$ assuming no slow reaction has occurred in $(t,t+\Delta t]$. If the residual $R_{j}(t)$ 
has performed a \textit{zero crossing} in $(t,t+\Delta t]$ then the $j$th slow reaction has occurred. We 
monitor $R_{j}(t)$ by writing equation (\ref{jump2}) in differential form,
\begin{equation}\label{jump3}
  \frac{dR_{j}(t)}{dt}=h_{j}^{s}(t,\bmc), \qquad R_{j}(t_{0})=\log(r_{j}).
\end{equation}
Equation (\ref{jump3}) can then be solved by using a time discretisation method such as the Euler scheme. 
Note that the method is restricted to only one slow reaction event in $(t,t+\Delta t]$. If more than 
one zero crossing occurs in this interval then $\Delta t$ can be reduced, and the state restored to the 
previous one. Hence, if the $j$ slow reaction occurs, the reaction time $\tau_{j}$ can be found through 
an It\^o-Taylor series expansion of (\ref{jump3}). If $t'$ is the time just prior to the $j$th slow reaction 
then
\[
\tau_{j}=-\frac{R_{j}(t')}{h_{j}^{s}(t',\bmc)}+t'.
\] 
The scheme provides an accurate way of capturing a slow reaction event provided that over the interval of interest, 
say $[t_{curr},t_{curr}+\Delta t_{integrate}]$, it is known that only one reaction occurs. Consequently, if more than 
one zero crossing is recorded, the interval length is reduced until at most one slow event is captured.  

The algorithm commences at time $t_{curr}=0$ with known rate constants $\bmc$, a known number molecules 
$\bmx_{curr}$ and $R_{j}(0)=\log(r_{j}),\, j=1,\ldots ,r^{s}$. The algorithm ends with $\bmx_{curr}$ 
as the state vector at time $t_{end}>t_{curr}$. For simplicity, we take the 
length of the time interval over which a slow reaction is detected to be $\Delta t_{integrate}= \Delta t_{hybrid}$.
\begin{enumerate}
\item \textbf{If} $t_{curr}\ge t_{end}$ then \textbf{stop}.
\item Set $\Delta t_{hybrid}=\min(\Delta  t_{hybrid},t_{end}-t_{curr})$.
\item {Classify reactions}: given $\bmx_{curr}$ classify each reaction as either slow or fast.
\item Calculate the fast reaction hazards. Using an Euler time step of $\Delta t_{euler}$, numerically integrate the SDE (\ref{daf}) for the fast 
  species over $(t_{curr},t_{curr}+\Delta t_{hybrid}]$ giving a sample path for the fast species over $(t_{curr},t_{curr}+\Delta t_{hybrid}]$.  
\item Using the slow reaction hazards, compute each residual $R_{j}(t)$, $j=1,\ldots ,r^{s}$ using an 
  Euler approximation of (\ref{jump3}) and decide whether or not a slow reaction has happened in $(t_{curr},t_{curr}+\Delta t_{hybrid}]$.
\item \textbf{If} {no slow reaction} has occurred, set $t_{curr}:=t_{curr}+\Delta t_{hybrid}$ and update the fast species to their proposed values 
  at $t_{curr}$; \textbf{go to Step 1}.
\item \textbf{If} {one slow reaction} has occurred, identify the type $j$ and time $\tau_{j}$, set $t_{curr}=\tau_{j}$ and update the 
  system to $\tau_{j}$ using the same random numbers as in step (d). Reset the $j$th residual, $R_{j}(t)=\log(r_{j})$. 
Reset $\Delta t_{hybrid}$ to its initial value if required. \textbf{Goto Step 1.}
\item \textbf{If} {more than one slow reaction} has occurred, reduce $\Delta t_{hybrid}$ and \textbf{goto Step 3}.
\end{enumerate}
Note that in step 3, for consistency, we use the same decision criteria outlined in Section~\ref{reactchoice}.


\bibliography{papers}	

\end{document}